# Chemical requirements for stabilizing type-II Weyl points in MnBi$_{2-x}$Sb$_x$Te$_4$

Yuanxi Wang

2-Dimensional Crystal Consortium, Materials Research Institute, The Pennsylvania State University, University Park, PA 16802, USA

**We show that type-II Weyl point formation in MnBi$_{2-x}$Sb$_x$Te$_4$ is more likely than in MnBi$_2$Te$_4$ when *x* reaches 0.5, as the alloy case does not suffer from the same degree of lattice parameter sensitivity as in MnBi$_2$Te$_4$. To further substantiate the stability of type-II Weyl points in MnBi$_{2-x}$Sb$_x$Te$_4$, we demonstrate that among the three conditions of establishing a type-II Weyl point, two are robustly satisfied by the zone-folded dispersion of Bi and Te $p_z$ orbitals and spin-orbit coupling already available in MnBi$_2$Te$_4$, and that the control over MnBi$_{2-x}$Sb$_x$Te$_4$ alloy composition provides a rational means to satisfy the third condition. The stability of type-II Weyl points in MnBi$_{1.5}$Sb$_{0.5}$Te$_4$ is thus intimately associated with orbital interactions, providing a concrete foundation for future efforts in band engineering and the rational design of topological electronic structures.**

Efforts to search for Weyl semimetals have recently concentrated on finding so-called *ideal* Weyl semimetals, where the number of Weyl cones in the Brillouin zone is at the minimum of two [1,2]. They are desired since unambiguous measurements of the topological signatures of Weyl semimetals rely on carefully aligning Fermi levels relative to the Weyl points; the fewer interruptions from other bands near the Fermi level the better. One general strategy of achieving ideal Weyl semimetals is to break only time-reversal symmetry while preserving inversion symmetry [1]; otherwise, the reverse – breaking only inversion symmetry while preserving time-reversal symmetry – would produce a minimum of four Weyl points. Such magnetic Weyl semimetals have been recently proposed [2,3] and experimentally investigated [4,5] in the intrinsic magnetic system MnBi$_2$Te$_4$ [6,7] in its ferromagnetic phase. The current theory challenge is that, at the density functional theory (DFT) level, the existence of type-II Weyl points in ferromagnetic-phase MnBi$_2$Te$_4$ is sensitive to lattice parameter changes on the order of 1% [2] (similar to sensitivities in WTe$_2$ [8,9]) and would seemingly rely on fine tuning to achieve. While this sensitivity is explicitly demonstrated by previous theory studies, achieving a type-II Weyl point *without* the requirement of fine-tuning lattice parameters is far from hopeless, and can be approached in two ways. The first is to demonstrate that one can expand the parameter space of the lattice parameter that achieves type-II Weyl points by tuning a second, well-defined, and controllable materials parameter. In the following we demonstrate that tuning $MnBi_{2-x}Sb_xTe_4$ alloy composition $x$ from 0 to 0.5 removes the requirement of artificial lattice dilation to achieve type-II Weyl points. The second approach is to articulate how exactly the conditions of realizing type-II Weyl points could be satisfied at all in $MnBi_2Te_4$ and how it becomes more likely in $MnBi_{1.5}Sb_{0.5}Te_4$ based on general principles of orbital interactions; this would establish the type-II Weyl point arising as an inevitability from a solid-state chemistry perspective, instead of relying on numerical results from DFT. Similar strategies based on chemical principles and heuristics (e.g. orbital interactions, band fillings, Peierls distortion, and zone-folding) and have already proven successful in predicting topological insulators [10] and topological semimetals [1,11,12]. Here we demonstrate that type-II Weyl cones in $MnBi_2Te_4$ originate from the zone folding of $p_z$ orbitals and the modification of band dispersions due to spin-orbit interaction.

We first demonstrate numerical results from DFT. The three panels in Fig. 1 show the low-energy DFT band structures for $MnBi_2Te_4$, $MnBi_2Te_4$ dilated by %1 in both the in-plane and out-of-plane directions, and $MnBi_{1.5}Sb_{0.5}Te_4$. All native lattice parameters are taken from Ref. [13], where $a$= 4.334 Å and $c$= 40.931 Å for $MnBi_2Te_4$, and $a$=4.310 Å and $c$ = 40.939 Å for $MnBi_{1.5}Sb_{0.5}Te_4$. For $MnBi_2Te_4$, the requirement of 1% dilation to achieve type-II Weyl points is consistent with Ref. [2]. For $MnBi_{1.5}Sb_{0.5}Te_4$, a type-II Weyl point is readily formed at its native lattice constant. Details on the electron and hole pockets in $MnBi_{1.5}Sb_{0.5}Te_4$ will be described elsewhere [5]. Out of the two bands forming the Weyl point, the band with the weaker dispersion obtains a more negative slope in $MnBi_{1.5}Sb_{0.5}Te_4$ than in 1%-dilated $MnBi_2Te_4$, even larger than in $MnBi_2Te_4$, indicating increased stability of the type-II-ness of the Weyl point. This is related to the smaller spin-orbit coupling (SOC) strength in Sb than in Bi, as elaborated later. Of course, systematic errors intrinsic to DFT may still affect the details of the band structures that determines whether the Weyl point survives; however, the above results indicate that achieving a type-II Weyl point in $MnBi_{1.5}Sb_{0.5}Te_4$ is at least more likely than in $MnBi_2Te_4$, all other things being equal.

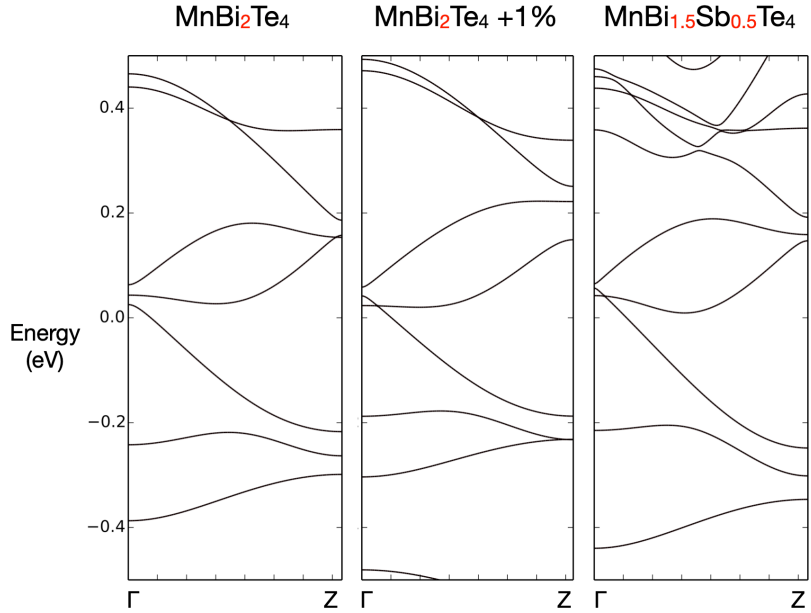

Figure 1. DFT band structures for MnBi$_2$Te$_4$ at its native lattice constant, MnBi$_2$Te$_4$ at 1% dilation, and MnBi$_{1.5}$Sb$_{0.5}$Te$_4$ at its native lattice constant. The 1% lattice dilation is required for MnBi$_2$Te$_4$ to achieve a type-II Weyl point, but not required for MnBi$_{1.5}$Sb$_{0.5}$Te$_4$.

We next investigate the origin of the type-II Weyl points in MnBi$_2$Te$_4$. By definition, type-II Weyl points require three conditions, two bands that disperses with the same sign in their slopes (type-II-ness) [14], with different magnitudes in their slopes (so that band crossings are possible at all), and with appropriate placements in energy (so that they do not miss each other). We show that the first two conditions are satisfied by zone-folded dispersions of Bi and Te $p_z$ orbitals and spin-orbit coupling effects already available in MnBi$_2$Te$_4$, and that the third condition (the only one that does require tuning) can be satisfied by tuning SOC strength, by means of tuning MnBi$_{2-x}$Sb$_x$Te$_4$ alloy composition.

It is well-known in solid-state chemistry literature that the electronic structure of heavy group 15 elements and their compounds with heavy group 16 elements (e.g. Bi$_2$Te$_3$) are mainly governed by the dispersion of $p_{x,y,z}$ orbitals [15–18]. In many cases, DFT band structures closely resemble ones from simple tight-binding models only considering $pp\sigma$ interactions [15,18]. MnBi$_2$Te$_4$ is similar, where the low-energy electronic structure is dominated by Bi and Te $p_{x,y,z}$ orbitals and where Mn effectively serves to donate its two 4$s$ electrons to the Bi and Te $p_{x,y,z}$ manifolds, with negligible hybridization near the Fermi level. One therefore anticipates that the band dispersion along $z$ (the out-of-plane direction) could be well-described by a 1D chain of $p_z$ orbitals with $pp\sigma$ interactions, i.e. a cosine-like dispersion starting from a band maximum at $\Gamma$ and a band minimum at the Brillouin zone edge, as shown by the black curve of Fig. 2a. Since a unit cell of $MnBi_2Te_4$ has six such $p_z$ orbitals (Fig. 2d), the dispersion along $z$ is then zone-folded six times as shown by the red curves in Fig.2a, forming a zigzagging $p_z$ dispersion that alternates six times between positive and negative slopes with increasing energy. This is exactly what is found in the DFT band structure of $MnBi_2Te_4$ without SOC in Fig. 2b. The deviations from the simple model is of course due to more intricate hybridization gap openings from Bi-Te interactions and further involvement of $s$ orbitals. The band gap in this non-SOC system lies between the fourth and fifth $p_z$ bands (see band labels in Fig. 2b), a feature that will become useful later when SOC is turned on. For the simple 1D model, we expect six states at $\Gamma$, the lowest-energy state being all bonding, the highest-energy state being all antibonding, and four intermediate states where the blocks of three upper orbitals and three lower orbitals are in-phase and out-of-phase respectively. This expectation again agrees qualitatively with the orbital diagram in Fig. 2e obtained from analyzing the crystal overlap Hamilton populations (COHP) [19] of states at $\Gamma$ calculated from DFT (see Supplemental Materials). The lowest-and highest-energy states are indeed all bonding and antibonding respectively. The four mid-energy states (split from the two doubly-degenerate states at $\Gamma$ in the 1D model) can be qualitatively described by two three-atom blocks that are in-phase or out-of-phase. This orbital diagram can also be understood by building up the MnBi$_2$Te$_4$ electronic structure element by element. Each MnBi$_2$Te$_4$ layer contains two outer Te, two inner Te, and two Bi atoms (Fig. 2d). Beginning with $p_z$, the two outer Te $p_z$ splits into bonding and antibonding (bands 2 and 4 in Fig. 2e), the two inner Te $p_z$ splits into bonding (band 1) and antibonding, where the latter further interacts with Bi antibonding $p_z$ and splits into band 3 and 6. The final Bi bonding $p_z$ is found at band 5. What remains in the blue bands in Fig. 2b are the $p_{x,y}$ orbitals interacting mainly through $pp\pi$ interactions, hence their much weaker dispersion than the $p_z$ states. Turning on spin-orbit coupling introduces energy splits of all the $p$ bands into $p_{3/2}$ and $p_{1/2}$ components as well as additional orbital mixings, while the overall electronic structure described above still holds, as shown in Fig. 2c. Judging from the orbital characters of the bands and the earlier observation that the Fermi level lies between the fourth and fifth $p_z$ band, we can still discern the six zigzagging $p_z$ bands, as highlighted in Fig. 2c.

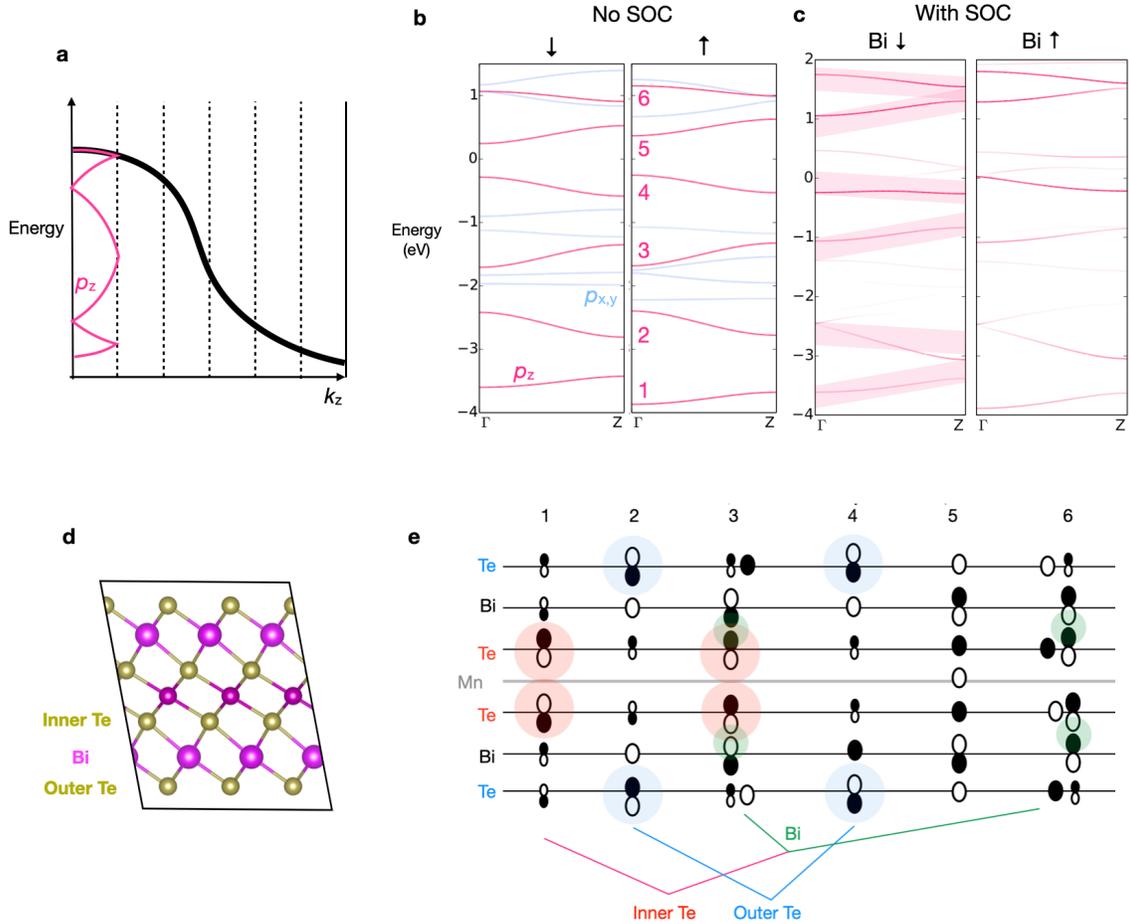

Figure 2. (a) 1D band dispersions derived from a minimal 1D chain model based on $p$ orbitals: original dispersion in black and zone-folded dispersion in red. (b) Actual DFT band structure of $MnBi_2Te_4$ without spin-orbit interactions. The two subpanels display two spin channels. (c) DFT band structure of $MnBi_2Te_4$ with spin-orbit interactions. The zigzag pattern of the $p_z$ bands remains discernable, as highlighted by the thick bars. (d) The septuple layered structure of $MnBi_2Te_4$. (e) Orbital diagram of the $p_z$ manifold of $MnBi_2Te_4$ electronic structure and how it can derived from Bi and Te orbital interactions. Each color appears in two states, one indicating bonding and other indicating antibonding.

We are now ready to analyze the low-energy states near the Fermi level and relate their behavior to type-II Weyl point formation. The MnBi$_2$Te$_4$ panel of Fig. 3 shows the MnBi$_2$Te$_4$ band structure in the energy range between –0.4 and 0.2 eV relative to the Fermi level, projected onto outer Te orbitals and Bi orbitals (left/right), and distinguishing between positive and negative $s_z$ projection (upper/lower). Projection onto $p_z$ and $p_{x,y}$ are indicated by red and blue respectively. This energy range contains four bands, labeled 1 through 4. We focus on the bands involved in shaping the Weyl point: band 2, the band with the weaker dispersion, band 3, the band with the stronger dispersion, and band 4, whose role will become clear later. Band 2 transitions from outer Te $p_z \downarrow$ to Bi $p_{x,y} \uparrow$ character (red to blue) going from $\Gamma$ to Z while band 4 makes the reverse transition: this pair is clearly the product of a Te $p_z \downarrow$ band and a Bi $p_{x,y} \uparrow$ band anticrossing due to SOC, consistent with SOC in $p$-orbital systems coupling $p_z$ with only $p_x$ or $p_y$ of the opposite spin. The unperturbed Te $p_z \downarrow$ band (i.e. before hybridizing with Bi $p_{x,y} \uparrow$) should therefore disperse towards lower energies from $\Gamma$ to Z: this is exactly what we expected earlier, where the fourth $p_z$ band in the non-SOC case is also of Te $p_z \downarrow$ character and disperses downwards. Its spin partner, the Te $p_z \uparrow$ band, is band 3 and fully expresses its downward dispersion without encountering anticrossings since it is well separated from Bi $p_{x,y} \downarrow$ in energy. Thus we conclude that bands 2 (near $\Gamma$) and band 3, the two bands involved in shaping the Weyl point, carry the same slope because they are both derived from the downward $p_z$ dispersion in the simple $p_z$-only model. The first condition for type-II Weyl point (same slope sign) is thus met. The SOC anticrossing encountered by band 2 weakens its dispersion compared with an unhybridized case, thus satisfying the second condition (different slope magnitude). In the DFT description of ferromagnetic-phase MnBi$_2$Te$_4$, the third condition is not met: the less dispersive band 2 lies *above* the more dispersive band 3, missing the opportunity to cross.

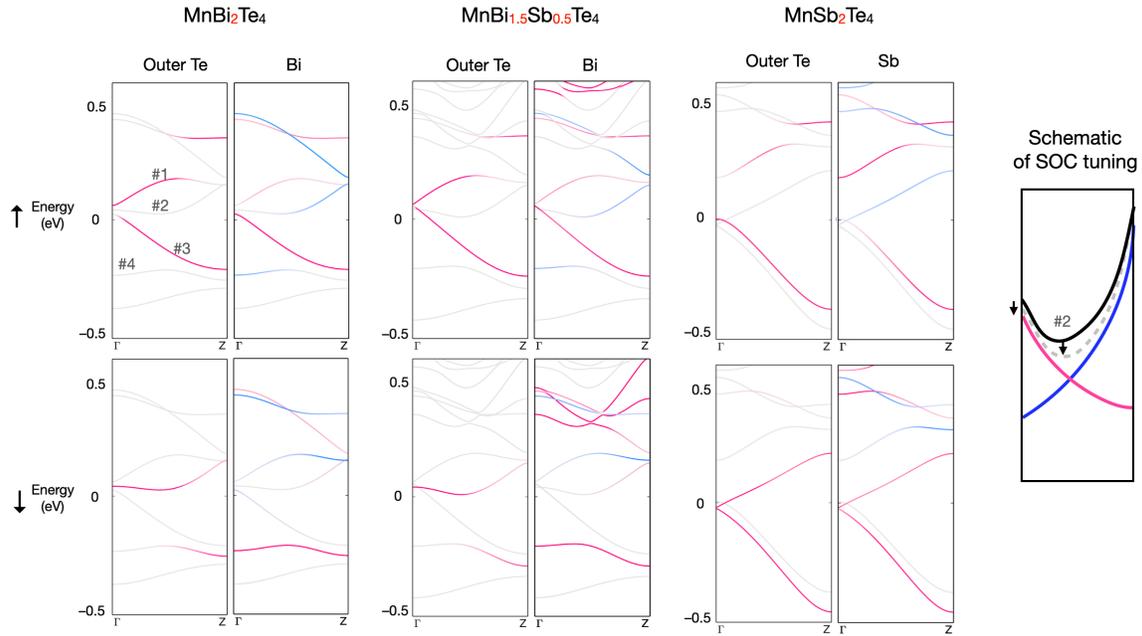

Figure 3. Projected band structures for $MnBi_2Te_4$, $MnBi_{1.5}Sb_{0.5}Te_4$, and $MnSb_2Te_4$. In each two-by-two panel, left/right subpanels are for projections onto outer Te and Bi orbitals and upper/lower subpanels are for projections onto positive and negative $s_z$.

So far we have established that the first two conditions are satisfied due to the underlying orbital chemistry. One can therefore narrow down the sensitivity of type-II Weyl point formation to the third condition, band ordering at $\Gamma$. Here we refrain from attempting to tune the band ordering for bands 2 and 3 at $\Gamma$ by applying strain, because it was already partially explored in previous work [2] and, more importantly, because strain simultaneously affects all orbitals, with energy shift directions that are not obvious *a priori*, while our goal is to limit the affected bands to bands 2 and 3, with the specific goal of inducing opposite energy shifts. What we investigate is substituting Bi with Sb in MnBi$_{2-x}$Sb$_x$Te$_4$ alloys, with the more predictable effect of reducing spin-orbit coupling strength, which will more strongly affect band 2 (since it is part of a SOC anticrossing pair). The result is already known from the previous section: bands 2 and 3 switch order at $\Gamma$ to form a type-II Weyl point at $x=0.5$. The reason band 2 lowers in energy at $\Gamma$ is that the reduced SOC brings band 2 closer to its pre-hybridization state, as schematically plotted in the rightmost panel of Fig. 3. In addition, the reduced SOC lowers the energy of band 2 at the anticrossing (midway between $\Gamma$ and Z), making the slope of the weakly dispersive band more negative and thereby stabilizing the type-II-ness of the Weyl point.

To further substantiate that band ordering at $\Gamma$ is mainly determined by SOC (instead of e.g. electronegativity differences between Bi and Sb), in Fig. 4 we show MnBi$_2$Te$_4$ band structures calculated with a different DFT implementation using the ABINIT package [20], where the local SOC can be artificially tuned from its full strength down to 60%. At full SOC strength, the band structure agrees with our previous MnBi$_2$Te$_4$ result, with minor quantitative differences. As SOC strength decreases from 100% to 80%, the gap between band 2 and 4 decreases, bringing band 2 to energies lower than band 3. Below 80% SOC strength, the Bi $p_{x,y}$ ↑ band gradually recovers its orbital character in a single band, dispersing upwards, and therefore crosses with the downward dispersing Te $p_z$ ↑ band to produce a type-I Weyl point. Meanwhile Te $p_z$ ↓ also recovers its orbital character in a single band. This MnBi$_2$Te$_4$ band structure at 80% SOC strength resembles that of MnSb$_2$Te$_4$ at its full SOC strength, also carrying a type-I Weyl point (Fig. 3, MnSb$_2$Te panel), suggesting that $x$ in MnBi$_{2-x}$Sb$_x$Te$_4$ beyond 0.5 will eventually convert the type-II Weyl point to type-I, attributed to the separation of Bi $p_{x,y}$ ↑ and Te $p_z$ ↓ as discussed above.

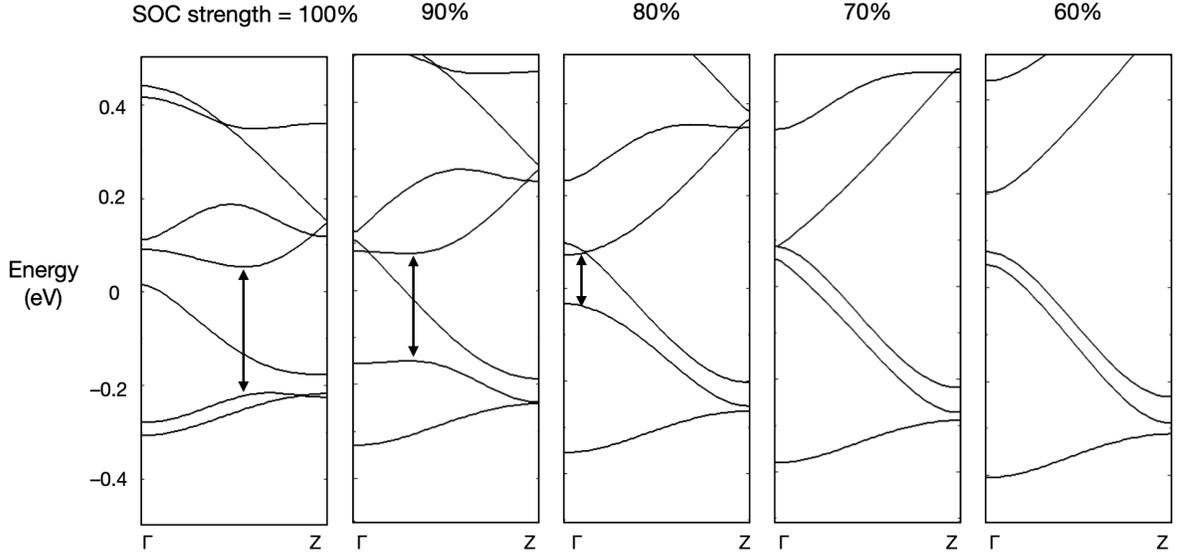

Figure 4. DFT band structures for MnBi$_2$Te$_4$, but with the local spin-orbit coupling tuned from its full strength down to 60%. The system evolves as follows with decreasing SOC strength: gapped (i.e. no band crossings near the Fermi level) → type-II Weyl semimetal → type-I Weyl semimetal → gapped.

In summary, we showed that type-II Weyl point formation in MnBi$_{1.5}$Sb$_{0.5}$Te$_4$ is more likely than in MnBi$_2$Te$_4$. We also demonstrate that among the three conditions of establishing a type-II Weyl point, two are robustly satisfied by the zone-folded dispersion of Bi and Te $p_z$ orbitals and spin-orbit coupling already available in MnBi$_2$Te$_4$, and that the third condition can be satisfied by tuning MnBi$_{2-x}$Sb$_x$Te$_4$ alloy composition. Compared with existing numerical studies, the formation and stability of type-II Weyl point in MnBi$_{1.5}$Sb$_{0.5}$Te$_4$ is thus placed on more concrete foundations based on orbital interactions.

**Methods**. Except for the adjustable SOC strength case, all density functional theory calculations were performed using the generalized gradient approximation exchange-correlation functional parametrized by Perdew, Burke, and Ernzerhof (GGA-PBE) [21,22], with plane-wave expansion cutoff energies of 500 eV and a Γ-centered k-point grid of 7×7×6, as implemented in the Vienna *Ab*-initio Simulation Package (VASP) [23]. Electron-ion interactions were described by projector augmented wave (PAW) pseudopotentials [24,25]. Structural relaxations were performed until forces were smaller than 0.01 eV/Å. On-site Coulomb interactions experienced by localized 3$d$ electrons in Mn were treated by the GGA+$U$ scheme, where (following Ref. [26]) we chose the effective $U$ parameters $U_{eff} = U - J = 4$ eV for 3$d$ electrons in Mn. Noncollinear magnetism and spin-orbit coupling were included using the native VASP implementation [27]. Van der Waals corrections were introduced by Grimme's semi-empirical DFT-D3 correction scheme [28]. The adjustable SOC strength case was performed using the ABINIT package [20] and PAW potentials from Refs. [29,30], and with all other convergence parameters being identical to those in the VASP calculations.

**Acknowledgements**. Funding for this work was provided by the 2D Crystal Consortium NSF Materials Innovation Platform under cooperative agreement DMR-1539916.